\documentclass[preprint,superscriptaddress]{revtex4-2}

\usepackage{chemformula} 
\usepackage[T1]{fontenc} 
\usepackage{amsmath}
\usepackage{amssymb}

\usepackage{siunitx}
\DeclareSIUnit\adu{ADU}
\DeclareSIUnit\electron{e^-}

\makeatletter

\renewcommand{\vec}[1]{\mathbf{#1}}
\newcommand{\uvec}[1]{\hat{\mathbf{#1}}}

\begin{document}

\author{Tamal Roy}
\thanks{Contributed equally}
\affiliation{Department of Mechanical and Process Engineering, ETH Zurich, Zurich, Switzerland}
\author{Peter T. Brown}
\thanks{Contributed equally}
\affiliation{Center for Biological Physics and Department of Physics, Arizona State University, Tempe, AZ, USA}
\author{Douglas P. Shepherd}
\email{douglas.shepherd@asu.edu}
\affiliation{Center for Biological Physics and Department of Physics, Arizona State University, Tempe, AZ, USA}
\author{Lisa V. Poulikakos}
\email{lpoulikakos@ucsd.edu}
\affiliation{Department of Mechanical and Aerospace Engineering, UC San Diego, La Jolla, CA, USA}

\title{Spatial wavefront shaping with a multipolar-resonant metasurface for structured illumination microscopy}

\begin{abstract}
Structured illumination microscopy (SIM) achieves superresolution in fluorescence imaging through patterned illumination and computational image reconstruction, yet current methods require bulky, costly modulation optics and high-precision optical alignment. This work demonstrates how nano-optical metasurfaces, rationally designed to tailor the optical wavefront at sub-wavelength dimensions, hold great potential as ultrathin, single-surface, all-optical wavefront modulators for SIM. We computationally demonstrate this principle with a multipolar-resonant metasurface composed of silicon nanostructures which generate versatile optical wavefronts in the far field upon variation of the polarization or angle of incident light. Algorithmic optimization is performed to identify the seven most suitable illumination patterns for SIM generated by the metasurface based on three key criteria. We find that multipolar-resonant metasurface SIM (mrm-SIM) achieves resolution comparable to conventional methods by applying the seven optimal metasurface-generated wavefronts to simulated fluorescent objects and reconstructing the objects using proximal gradient descent. The work presented here paves the way for a metasurface-enabled experimental simplification of structured illumination microscopy.
\end{abstract}

\maketitle

\section*{Introduction}

Structured illumination microscopy (SIM) is a fluorescence superresolution microscopy technique which can reach twice the diffraction limit by exploiting patterned excitation light combined with spatial frequency downmixing that occurs in incoherent imaging \cite{saxena2015structured}. SIM is increasingly popular because it is compatible with live-cell imaging, does not require specific fluorophore photophysics, and can capture images at rates approaching standard widefield microscopy~\cite{Heintzmann1999, Gustafsson2000, Prakash2021}. Most modern SIM instruments rely on illumination patterning optics, typically diffraction gratings or spatial light modulators (SLM's), placed in a conjugate imaging plane. Consequently, these experiments need precise optical alignment~\cite{Ball2015, Demmerle2017}, require considerable optical expertise, and are expensive to build.

Recently, a variety of approaches have emerged to simplify, democratize, and improve the speed,  stability, and efficiency of SIM experiments. These include fiber-based~\cite{hinsdale_high-speed_2021, calvarese_integrated_2022}, digital micromirror device (DMD)-based~\cite{Sandmeyer2021, Brown2021, Li2020a}, and photonic chip-based~\cite{Helle2020, lin_uv_2023} SIM schemes. Such efforts compliment parallel computational imaging approaches to limit reconstruction artifacts and achieve high-quality imaging at low signal-to-noise ratios~\cite{Huang2018, Qiao2022}. While each of these approaches is promising, none fully solves the problems discussed above. For example, fiber-based approaches simplify alignment but require expensive or complex methods to phase shifting such as Pockels cells or thermal phase-shifters. DMD-based approaches significantly reduce the cost of SIM setups but the diffraction physics of the DMD introduces new complexity and places constraints on multicolor operation. Photonic chip approaches offer the most compact and simplified systems, but are either limited to near-field excitations or have only been demonstrated for 1D resolution enhancement. 

Progress in nanoscale optics has enabled emerging technologies for generating structured light, which offer solutions to the abovementioned challenges. Nano-optical metasurfaces are composed of sub-wavelength-periodic optical elements that control light--matter interactions at the nanoscale. The high-spatial-frequency evanescent near fields generated by plasmonic metasurfaces -- those composed of appropriate metallic nanostructures \cite{maier2007plasmonics} -- have been applied to SIM to enhance spatial resolution~\cite{Wei2010, bezryadinaLocalizedPlasmonicStructured2017, ponsettoExperimentalDemonstrationLocalized2017, bezryadinaHighSpatiotemporalResolution2018}. However, such resolution enhancement can only be achieved at nm-scale proximity to the metasurface, thus necessitating close contact to the sample of interest and hindering applications such as depth-dependent confocal imaging. Moreover, this method still requires complex piezoelectric stages to achieve illumination patterns, maintaining the experimental complexity of conventional SIM experiments. Beyond near-field effects, metasurfaces enable versatile avenues for light manipulation in the far field.~\cite{yu2011light} Metasurfaces have been rationally designed for full phase and polarization control of incident light,~\cite{arbabi2015dielectric} leading to numerous applications encompassing flat optical elements,\cite{lin2014dielectric, lindfors2016imaging, khorasaninejad2016metalenses, klopfer2020dynamic} beam steering,\cite{holsteen2019temporal, thureja2020array, lawrence2020high} holography,\cite{huang2013three, ni2013metasurface, zheng2015metasurface} or optical computation,\cite{silva2014performing, mohammadi2019inverse, cordaro2023solving} as has been reviewed in prior work.~\cite{yu2014flat, genevet2017recent, kamali2018review, shaltout2019spatiotemporal, barton2020high, thureja2022toward} 

Notably, metasurfaces have unique capabilities to control the optical wavefront at subwavelength dimensions through engineered manipulation of the phase of incident light \cite{arbabi2015dielectric, hail2022high}. However, the potential of metasurfaces to act as ultrathin, single-surface, all-optical, generators of tunable SIM illumination patterns, remains largely unexplored. Wavefront-shaping capabilities of metasurfaces present an exciting opportunity to drastically simplify the experimental configuration of SIM by leveraging the ability of metasurfaces to generate far-field patterns in the optical wavefront. Moreover, all-optical tunability of the metasurface-generated wavefronts are enabled by tailoring the interaction of the metasurface with, e.g., varying polarization states or incident angles of the illumination source. Superresolution images can then be obtained by tailored computational reconstruction algorithms. 

In this work we introduce multipolar-resonant metasurface SIM (mrm-SIM), an approach which retains the compact system found in photonic chip-based approaches but uses far-field patterns and achieves 2D resolution enhancement. mrm-SIM generates structured light by resonant interactions with nano-optical structures and allows the illumination patterns to be tuned by changing the incident polarization and angle. With mrm-SIM we achieve resolutions up to \qty{240}{\nano \meter} for \qty{1004}{\nano \meter} excitations, comparable to those achieved with conventional hexagonal SIM. By embedding such a device in a glass coverslip we propose a nearly alignment-free SIM approach. Since mrm-SIM uses far-field patterns, we anticipate it can be extended to 3D SIM by appropriate pattern engineering. Furthermore, unlike most other SIM approaches the achievable field-of-view (FOV) in this approach is limited by the size of the metasurface instead of the FOV of the detection system. Therefore, we expect mrm-SIM to be more scaleable and achieve larger spatial-bandwidth products than other approaches. This research paves the way toward significantly simplifying the experimental configuration required for SIM and the democratization and broad dissemination of superresolution microscopy techniques.

\section*{Results and discussion}
\begin{figure}[htb!]
	\centering
	\includegraphics[width=\linewidth]{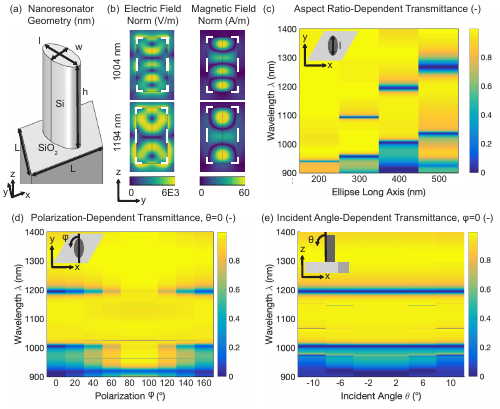}
	\caption{
    \textbf{Nanoresonator characterization.}(a) Nanoresonator design and geometry composed of a \ch{Si} nanoresonator on a \ch{SiO2} substrate in air. The nanoresonator has a height \textit{h} = \qty{715}{\nano \meter} and an elliptical cross sectional area with width \textit{w} = \qty{200}{\nano \meter} and length \textit{l} = \qty{400}{\nano \meter} and is arranged periodically in an equilateral rhombohedral lattice with side length \textit{L} = \qty{670}{\nano \meter}. (b) Electric (left) and magnetic (right) field norms of the nanoresonator at resonant wavelengths of $\lambda$ = \qty{1004}{\nano \meter} (top) and $\lambda$ = \qty{1194}{\nano \meter} (bottom), indicating both electric and magnetic multipolar resonant behavior. The incident plane wave intensity is \qty{1000}{\watt / \meter^2}. (c-e) Transmittance spectra for the nanoresonator for varying parameters. (c) Varying long \textit{l} axis of elliptical cross section for $\theta$ = 0, $\varphi$ = 0, $w = \qty{200}{\nano \meter}$. (d) Varying polarization $\varphi$ for \textit l = \qty{400}{\nano \meter}, $w = \qty{200}{\nano \meter}$, $\theta$ = 0. (e) Varying incident angle $\theta$ for $\varphi$ = 0, \textit{l} = \qty{400}{\nano \meter}, $w = \qty{200}{\nano \meter}$.
    \label{fig:nanostructure_resonance}
    }
\end{figure}

The wavefront-shaping metasurface presented in this work is based on multipolar-resonant silicon nanoresonators,\cite{schuller2007dielectric, van2013designing, evlyukhin2013multipole, miroshnichenko2015substrate, butakov2016designing, hail2022high} with elliptical cross sections of short axis \textit{w}, long axis \textit{l} and height \textit{h}. For initial characterization, the nanostructures are arranged in an equilateral rhombohedral lattice with side length \textit{L} = \qty{670}{\nano \meter} (Fig.~\ref{fig:nanostructure_resonance}a) and periodic boundary conditions. For our purposes, effective multipolar wavefront shaping is achieved at infrared frequencies where silicon absorption becomes negligible and electric and magnetic multipolar resonances overlap, thus enabling total transmission blocking at resonant frequencies.\cite{hail2022high} For a nanostructure of \textit{w} = \qty{200}{\nano \meter}, \textit{l} = \qty{400}{\nano \meter} and \textit{h} = \qty{715}{\nano \meter}, Fig.~\ref{fig:nanostructure_resonance}b shows cross sectional near-field plots of the electric and magnetic field norms, demonstrating the simultaneous occurrence of electric and magnetic multipolar resonances at \qty{1004}{\nano \meter} and \qty{1194}{\nano \meter} upon $y$-polarized incident light.

The aspect ratio of the elliptical cross sectional area dictates the resonant behavior and optical tunability of the metasurface response. Figure~\ref{fig:nanostructure_resonance}c demonstrates variation of the nanoresonator aspect ratio by changing the long axis \textit{l} of the structure from a circular to elliptical cross sectional area in \qty{100}{\nano \meter} steps. Based on this parameter study, we select an ellipse with \textit{w} = \qty{200}{\nano \meter} and \textit{l} = \qty{400}{\nano \meter} for further investigation, as the multipolar resonances along the long and short ellipse axes will be spaced sufficiently far apart to ensure their independent behavior, while maintaining narrow-bandwidth resonant properties.

The asymmetrical elliptical nanoresonator cross section and the rhombohedral unit cell of the periodic boundaries result in an optically tunable metasurface response. Figure 1d,e demonstrates how varying the polarization ($\varphi$) and incoming angle ($\theta$) can controllably tune the transmittance at a given wavelength of interest. We define the polarization angle $\varphi$ from the $\uvec{y}$ axis, i.e. the electric field points along $\uvec{\epsilon} = - \sin \varphi \uvec{x} + \cos \varphi \uvec{y}$ and the incidence angle is in the $xz$ plane so that the beam propagates along $\uvec{k} = -\sin \theta \uvec{x} + \cos \theta \uvec{z}$. Figure~\ref{fig:nanostructure_resonance}d ($0^\circ\leq \varphi \leq 160^\circ$ with  $20^\circ$ increments, $\theta$ = 0$^\circ$), shows that varying the incident polarization state $\varphi$ leads to complete tunability of the transmittance at the multipolar resonant wavelengths $\lambda = \qty{1004}{\nano \meter}$ and $\lambda = \qty{1194}{\nano \meter}$. In particular, transmittance is completely blocked for excitation along the long axis of the ellipse ($\varphi = 0^\circ$). Figure~\ref{fig:nanostructure_resonance}e ($\varphi$=0$^\circ$, $-10^\circ \leq \theta \leq 10^\circ$ with  $2^\circ$ increments), demonstrates that varying the incident angle $\theta$ does not significantly alter the metasurface multipolar resonant transmittance. The optical properties observed in Fig.~\ref{fig:nanostructure_resonance}d,e indicate all-optical tunability of the studied multipolar resonant metasurface and will now be leveraged to rationally design a wavefront-shaping metasurface for SIM.

\begin{figure}[htb!]
	\centering
	\includegraphics[width=0.7\linewidth]{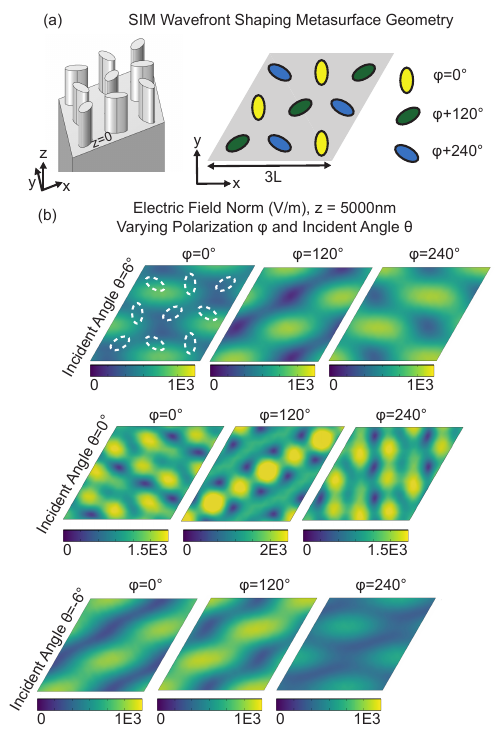}
	\caption{
    \textbf{Design and characterization of wavefront shaping metasurface for SIM.} (a) Metasurface geometry is composed of an equilateral rhombohedral superlattice of 9 cells with side length 3$L$ = \qty{2010}{\nano \meter} (left). Each cell contains a nanoresonator as described in Fig.~\ref{fig:nanostructure_resonance} at rotations of \qty{0}{\degree}, \qty{120}{\degree} and \qty{240}{\degree} (right). (b) Electric field norm of the far-field patterns generated by the metasurface at z = \qty{5000}{\nano \meter} for varying incident angles of $\theta = \{ -6^\circ, 0^\circ, 6^\circ \}$ (bottom to top) and polarization states $\varphi = \{ 0^\circ, 120^\circ, 240^\circ \}$.}
	
\label{fig:sim_metasurface}
\end{figure}

To generate optically tunable far-field wavefronts which are suitable for SIM, we designed a metasurface where the elliptical nanostructures described in Fig.~\ref{fig:nanostructure_resonance} are arranged in a periodically repeating geometry which generates intensity patterns with high-contrast and large spatial frequency support. These patterns are formed by constructive and destructive interference of light from the multipolar nanoresonators characterized in Fig.~\ref{fig:nanostructure_resonance}. Using multipolar resonances ensures the patterns have regions of low intensity, and hence high contrast. The lattice geometry with rotational symmetry generates the large spatial frequencies and ensures coverage in different spatial directions, which is vital for achieving isotropic resolution enhancement. SIM requires a diversity of different intensity patterns, and these can be generated by varying the incident angle $\theta$ and polarization angle $\varphi$ while retaining the high contrast and high spatial frequency content imposed by the metasurface. Therefore, our metasurface generates all-optically tunable, SIM-appropriate wavefront patterns.

Figure~\ref{fig:sim_metasurface}a shows a three-dimensional rendering (left) and top-down schematic (right) of our SIM wavefront shaping metasurface. The metasurface consists of a periodically arranged equilateral rhombohedral supercell with side length 3\textit{L} = \qty{2010}{\nano \meter}. The supercell contains 9 nanoresonators, with dimensions described in Fig.~\ref{fig:nanostructure_resonance}, each located at the center of an equilateral rhombohedral sub-unit-cell with side length \textit{L} = \qty{670}{\nano \meter}. As shown in the right panel, the nanoresonators are arranged at varying rotations of \qty{0}{\degree} (yellow), \qty{120}{\degree} (green) and \qty{240}{\degree} (blue), respectively. Due to the rotational symmetry of the nanoresonator arrangement, varying polarization states $\varphi$ will result in varying transmittance values for each nanoresonator. For example for $\varphi = \{0^\circ, 120^\circ, 240^\circ\}$, the nanoresonators marked in yellow, green or blue, respectively, will be excited along their long axis, resulting in transmittance blocking (see Fig.~\ref{fig:nanostructure_resonance}d). Moreover, varying incident angles $\theta$ do not significantly alter the multipolar resonant wavelength (Fig.~\ref{fig:nanostructure_resonance}e), yet varying $\theta$ will alter the propagation of transmitted light to the far field for mutually rotated nanoresonators. This polarization and angle-dependent excitation of the multipolar nanoresonators enables the generation of versatile, SIM-appropriate wavefront patterns in the far field through constructive and destructive superposition of the optical response of each nanoresonator.

Figure~\ref{fig:sim_metasurface}b demonstrates examples of far field wavefront patterns generated by the SIM wavefront shaping metasurface at height $z = \qty{5000}{\nano \meter}$ from the metasurface substrate for varying incident polarizations $\varphi = \{0^\circ, 120^\circ, 240^\circ\}$ and angles $\theta = \{-6^\circ, 0^\circ, 6^\circ\}$ (see Figures \ref{fig:S1} and \ref{fig:S2} in the Supporting Information for further far-field analysis). Dashed lines in the top left panel indicate the position of the nanoresonators at $z = 0$. These examples demonstrate the suitability of the generated wavefronts for SIM as they obey the above-noted criteria of high contrast and large spatial frequency support. Moreover, they exhibit similar wavefront patterns for a given $\theta$ which rotate and translate at varying $\varphi$.

We can better understand the structure of the mrm-SIM patterns by considering the underlying symmetry of the device. The metasurface consists of a periodic array of pillars arranged on a hexagonal (triangular) lattice with primitive lattice vectors $\vec{a}_1 = \qty{670}{\nano \meter} \times (3, 0)$ and $\vec{a}_2 = \qty{670}{\nano \meter} \times (3 \cos \qty{60}{\degree}, \sin \qty{60}{\degree})$, together with a three element basis describing the position of the three non-equivalent posts. The underlying lattice has $D_6$ dihedral symmetry while the basis has $D_3$ dihedral symmetry. The rhombohedral supercell the simulations are performed on contains three primitive cells (Fig.~\ref{fig:sim_metasurface}a). This supercell has less symmetry than the full lattice, as its only symmetry is a reflection plane \qty{30}{\degree} counterclockwise from the $x$-axis. This symmetry guarantees that the patterns for $\varphi = \qty{0}{\degree}, \qty{240}{\degree}$ polarization are reflected copies of each other for $\theta = \qty{0}{\degree}$ as illustrated in Fig.~\ref{fig:sim_metasurface}b because these polarization directions are \qty{\pm 60}{\degree} from the reflection axis.

\begin{figure}[htb!]
	\centering
	\includegraphics[width=\linewidth]{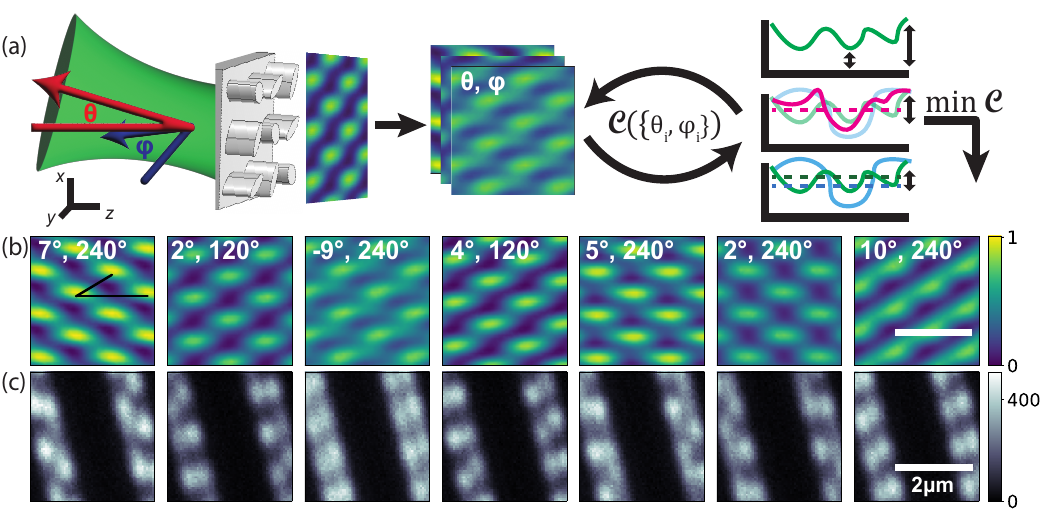}
	\caption{
    \textbf{SIM pattern optimization}.
	(a) We select \num{7} patterns from an initial set of \num{21} incidence angles $\theta$ and \num{3} polarization angles $\varphi$ by choosing a pattern set which optimizes the individual pattern contrast (top right), flat sum intensity for the pattern set (center right), and intensity balance between the various patterns (bottom right). Here the green and blue curves are cartoons of 1D intensity line cuts of individual patterns, and magenta of the sum pattern.
    (b) An optimal set of \num{7} intensity patterns and the incident angles and polarization angles $\theta, \varphi$. Lattice vectors are shown in black (left, inset).
    (c) Simulated images of the patterns from part b applied to a set of fluorescent line pairs with variable spacing. The number of photons for each pixel is indicated by the color scale (right panel).
\label{fig:pattern_optimization}}
\end{figure}

Once the metasurface geometry is fixed, we select a set of illumination patterns which are suitable for superresolution SIM from the range of polarization and incidence angles simulated previously. We assume the electric field above the metasurface is relayed to the sample, and defer discussion of proposed experimental geometries. To choose an optimal set of patterns, we developed an algorithm to select $N$ patterns which have similar characteristics to the classical patterns used in sinusoidal and hexagonal SIM.  Our algorithm, which is illustrated in Fig.~\ref{fig:pattern_optimization}a, relies on the following heuristic features of high-quality SIM patterns: (i) each individual pattern should have high contrast, with regions where there is no optical intensity (ii) the sum intensity of all the patterns should be spatially uniform (iii) all patterns should have the same peak intensity (iv) all patterns should have large spatial-frequency support. To quantify these features, we define a cost function which formalizes the first three requirements as separate terms
\begin{equation}
\mathcal{C} \left( \{I_i \}_{i=1}^N \right) = \frac{\text{sd}(\bar I)}{\bar I} +  \frac{1}{N}\sum_i \frac{\min I_i}{\max I_i} + \frac{1}{N} \sum_i \left(1 - \frac{\max I_i}{I_{\max}} \right). \label{eq:pattern_cost_function}
\end{equation}
Here $I_i$ is the $i$th intensity pattern and $\bar{I}$ and $I_{\max}$ are the the average and maximum of the $N$ intensity patterns respectively. While requirement (iv) could also be included in the cost function, we find all patterns considered here have high-spatial frequency content so this is not necessary. For $M$ simulated patterns there are $\binom{M}{N}$ pattern sets, which is too large a space to exhaustively search. We randomly sample \num{150000} combinations and choose the pattern set with the lowest cost. Then, we take this optimal set and iteratively replace each pattern by any patterns not contained in the set. If this lowers the cost, we retain the new set of patterns.

The structure of our mrm-SIM patterns more closely resemble the hexagonal SIM patterns than they do sinusoidal SIM patterns. To perform a fair comparison with standard hexagonal SIM we chose $N = \num{7}$ and set the maximum spatial frequency of the hexagonal SIM patterns to match that in the mrm-SIM patterns (section~\ref{section:hex_sim}). The selected patterns, shown in Fig.~\ref{fig:pattern_optimization}b, fulfill requirements (i)--(iv). The total cost from eq.~\ref{eq:pattern_cost_function} is \num{0.36} with (i)--(iii) contributing \num{0.14}, \num{0.09}, and \num{0.13} respectively. We display the images which result from applying these patterns to a synthetic fluorescent object in Fig.~\ref{fig:pattern_optimization}c. 

\begin{figure}[htb!]
	\centering
	\includegraphics[width=\linewidth]{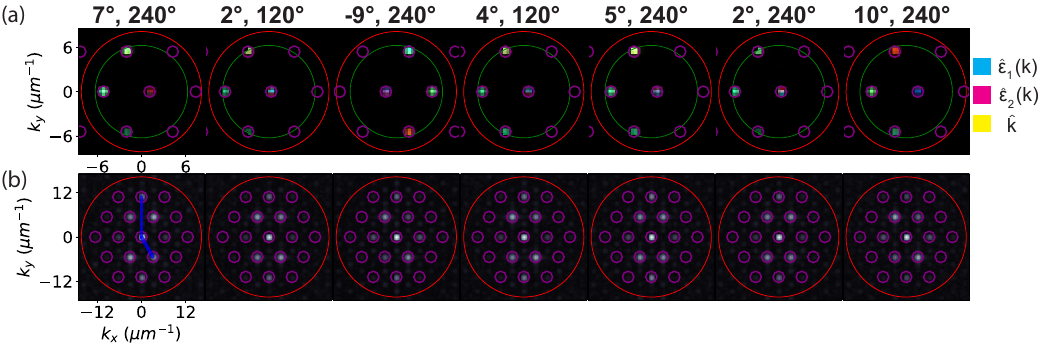}
	\caption{
    \textbf{SIM pattern spatial Fourier structure}.
    (a) Amplitude of the Fourier transform of the phasor electric fields corresponding to the intensity patterns shown in Fig.~\ref{fig:pattern_optimization}b. The maximum free-space propagating frequency (green), microscope electric field band pass frequency (red), and vectors $k_\perp(n_1, n_2)$ (purple) are shown. The electric field magnitude is illustrated using the color scale, $\uvec{\epsilon}(\vec{k})_{1,2}$ (cyan, magenta) and $\uvec{k}$ (yellow).
    (b) Amplitude of the Fourier transform of the intensity patterns shown in Fig.~\ref{fig:pattern_optimization}b (gamma = \num{0.3}). The (intensity) band pass frequency of the microscope is shown in red. The reciprocal lattice basis vectors are shown in blue, and other reciprocal lattice frequencies within the band pass frequency are shown in purple. The mrm-SIM patterns are primarily composed of \num{9} Fourier modes.
    \label{fig:sim_fourier_analysis}
    }
\end{figure}

We can better understand the spatial structure of the mrm-SIM patterns by exploiting the periodic structure of the metasurface. Due to Bloch's theorem, the spatial frequency content of the mrm-SIM electric field patterns is constrained by the primitive reciprocal lattice vectors, defined such that $\vec{a}_i \cdot \vec{b}_j = \delta_{ij}$. In this case, $\vec{b}_1 = (1/3, -1/\sqrt{3}) \times 1 / \qty{670}{\nano \meter}$ and $\vec{b}_2 = (0, 2/\sqrt{3}) \times 1 / \qty{670}{\nano \meter}$. A plane wave incident through the glass with wavevector $\vec{k} = k_\perp^\text{in} + k_z \uvec{z}$, where $|\vec{k}| = 2\pi n/\lambda$, interacts with the metasurface and couples into various modes which have 2D wavevectors
\begin{equation}
k_\perp(n_1, n_2) = k_\perp^\text{in} + 2\pi \left( n_1 \vec{b}_1 + n_2 \vec{b}_2 \right)
\end{equation}
for $n_1, n_2 \in \mathbb{Z}$. On the other side of the metasurface, the light is diffracted into the set of plane waves with wavevectors $\vec{k} = k_\perp(n_1, n_2) + \sqrt{\left(2\pi/\lambda \right)^2 - |k_\perp|^2} \uvec{z}$ and polarizations $\uvec{\epsilon}_2(\vec{k}) = \uvec{k} \times \uvec{z} / \left \vert \uvec{k} \times \uvec{z} \right \vert$ and $\uvec{\epsilon}_1(\vec{k}) = \uvec{\epsilon}_2(\vec{k}) \times \uvec{k}$. Where $|k_\perp(n_1, n_2)| \leq k$, these modes are propagating plane waves which describe the far-field intensity pattern produced by the metasurface. The intensity of the light that couples into each plane wave is controlled by the physics of the light--metasurface interaction and varies with incident angle and polarization. 

We first consider the electric field structure produced by the metasurface for orthogonal excitation ($\theta = \qty{0}{\degree}$). Due to our choice of reciprocal lattice vectors, only the shortest \num{6} non-zero reciprocal lattice vectors correspond with propagating modes in vacuum. In particular, $\vec{b_2} \lambda \sim \num{0.999}$, and there are \num{5} other lattice vectors of the same length due to the lattice symmetry. Similar to a Bessel beam~\cite{cespedesvicenteBesselBeamsUnified2021}, the plane-wave components at the symmetric spatial frequencies experience the same phase evolution as they propagate and hence they form an invariant pattern. The only pattern variation with $z$-position comes from interference between this fixed pattern and the DC term. Introducing a finite-sized beam will broaden these peaks and lead to additional divergence and pattern evolution.

When a tilt angle is introduced, this shifts the supported frequency by $k_\perp^\text{in} = -(2\pi n / \lambda) \sin \theta \uvec{x}$. Due to the aggressive choice of reciprocal lattice vectors even a \qty{1}{\degree} tilt converts one of the two supported reciprocal lattice vectors along the $x$-axis into an evanescent wave. This effect is clear in the Fourier transform of the phasor electric field shown illustrated in Fig.~\ref{fig:sim_fourier_analysis}a. For small tilts, the spatial frequencies maintain approximately the same magnitude and hence the patterns evolve little over a significant $z$ propagation distance (see section~\ref{section:far_field}). To display these Fourier peaks clearly we apodize the real-space pattern with a 2D Hann window and Fourier broaden them by considering a \qtyproduct{\sim 15 x 15}{\micro \meter} (\numproduct{500 x 500} pixel) area.

The structure of the mrm-SIM intensity patterns considered in Fig.~\ref{fig:pattern_optimization}b,c can be understood from similar considerations. The Fourier transform of the intensity pattern is related to the auto-correlation of the Fourier transform of the electric field components, hence all supported spatial frequencies are the difference of two electric field spatial frequencies, i.e. $\vec{k}_\text{int} = \vec{k}_\perp(n_1, n_2) - \vec{k}_\perp(m_1, m_2)$ which are necessarily reciprocal lattice frequencies. For the intensity patterns discussed above, we find only \num{9} peaks out of the \num{19} reciprocal lattice vectors within the band pass contribute strongly (Fig.~\ref{fig:sim_fourier_analysis}b), as these are the only peaks formed by auto-correlation of the \num{4} peaks in Fig.~\ref{fig:sim_fourier_analysis}a. The relative weight of the various Fourier peaks is shown in Fig.~\ref{fig:sim_fourier_analysis}b. The most important peaks for frequency resolution enhancement in SIM are the $\pm \vec{b}_2$ peaks found at normalized frequency \num{\approx 0.66} and $\pm(2 \vec{b}_1 + \vec{b}_2)$ at \num{\approx 0.38}. The absence of modes at the highest spatial frequencies along directions other than $\uvec{y}$ is due to the beam tilt effect discussed above.

\begin{figure}[htb!]
	\centering
	\includegraphics[width=\linewidth]{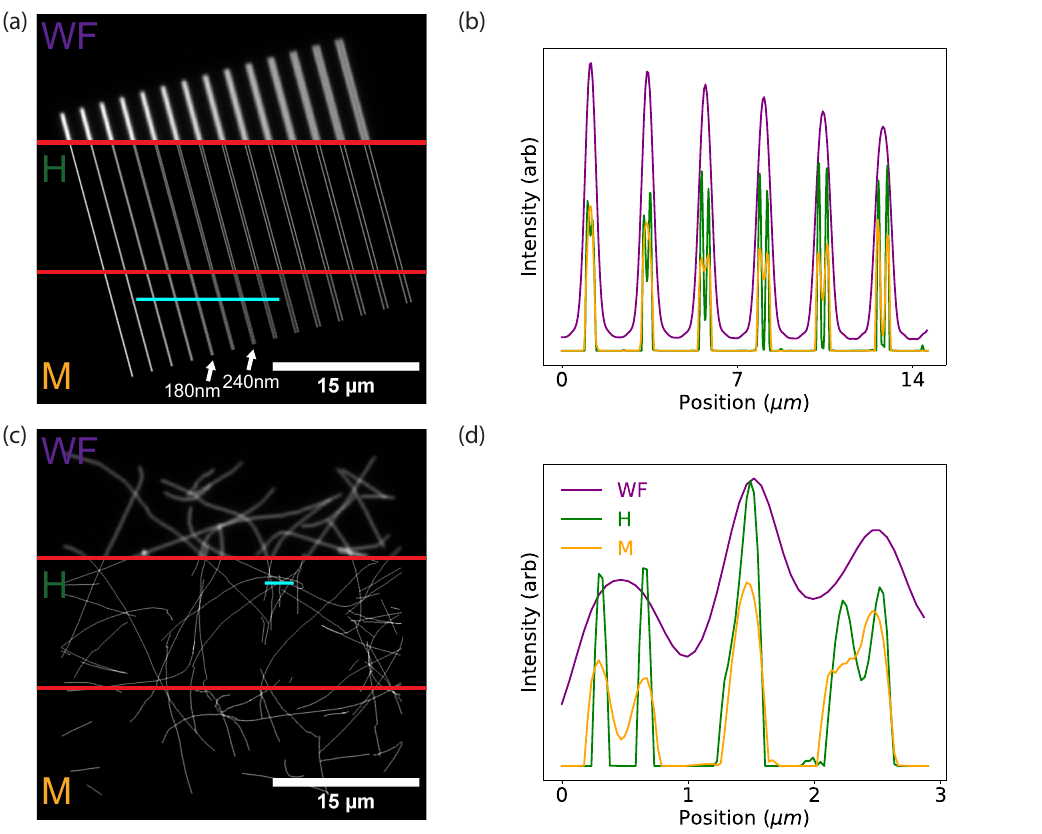}
	\caption{
    \textbf{SIM reconstruction}.
	(a) Widefield (WF), hexagonal SIM reconstruction (H), and mrm-SIM reconstruction (M) for sets of line pairs with spacings of \qtyrange{60}{450}{\nano \meter} in \qty{30}{\nano \meter} steps. Arrows indicate the closest resolvable line pairs which are \qty{180}{\nano \meter} for hexagonal SIM (left arrow) and \qty{240}{\nano \meter} for mrm-SIM. 
    (b) Widefield (purple), hexagonal SIM reconstruction (green), and mrm-SIM reconstruction (orange) line cuts along the path in part a (cyan line). This shows line pairs with spacings of \qtyrange{180}{330}{\nano \meter}.
    (c) Widefield, hexagonal SIM, and mrm-SIM reconstructions for synthetic filaments.
    (d) Line cuts through the synthetic filament images along the path shown in part c (cyan lines). 
    \label{fig:sim_reconstruction}
    }
\end{figure}

To quantify the quality of our mrm-SIM approach, we simulated images of two types of synthetic fluorescent samples: simulated line pairs with variable spacing and simulated filamentous structures (see section~\ref{section:simulating_imgs} and \ref{section:synthetic_objects}). We then reconstructed the underlying fluorescent object using an iterative reconstruction approach described below. In all cases we assume a numerical aperture of 1.3 and consider images with up to \num{1000} photons per pixel, a reasonable signal level for biological samples.

Due to the complexity of our patterns, applying standard SIM reconstruction approaches cannot effectively infer superresolution information. Instead, we adopt an iterative optimization-based reconstruction approach which relies on our prior knowledge of the pattern structure. This approach infers the sample structure using the fast iterative shrinkage-thresholding algorithm (FISTA)~\cite{Beck2009}, an accelerated proximal gradient-descent approach which incorporates constraints and regularization (see section~\ref{section:sim_recon_supplement}). In particular, we apply a positivity constraint and use total-variation regularization to enforce smoothness. A similar approach was recently applied to 3D SIM~\cite{Cai2022}. For realistic instruments, blind-SIM style pattern recovery may be employed to optimize the patterns~\cite{jostOpticalSectioningHigh2015}.

We find mrm-SIM patterns produce high quality reconstructions of the synthetic line pair images, as shown in Fig.~\ref{fig:sim_reconstruction}a. For the line pairs we chose spacings of \qtyrange{60}{450}{\nano \meter} in steps of \qty{30}{\nano \meter}. mrm-SIM resolves the line pair separated by \qty{240}{\nano \meter}, while the hexagonal SIM resolves the line pair separated by \qty{180}{\nano \meter}. Here we rely on the Sparrow criterion. Both resolution values should be compared with the maximum expected frequency enhancement in SIM, which is $1 + \frac{f_\text{sim}}{2 \text{NA} / \lambda}$ based on the maximum spatial frequency imaged by the microscope. For the patterns chosen here this is $\approx 1.67$, which corresponds to a resolution of \qty{230}{\nano \meter}. We also compare the widefield, hexagonal SIM reconstructed object and mrm-SIM reconstructed object by taking a line cut (Fig.~\ref{fig:sim_reconstruction}b). mrm-SIM clearly resolves line pairs which are not resolved in the widefield image. However, it does not achieve as high resolution or contrast as hexagonal SIM. We also considered a more realistic model of a biological sample: simulated filament networks modelling microtubules (Fig.~\ref{fig:sim_reconstruction}c,d). This simulation shows similar results to the synthetic line-pairs. 

The modest difference in the resolution and contrast of mrm-SIM and hexagonal SIM reconstructions can be understood by considering the distribution of pattern weight at different spatial frequencies. The hexagonal SIM patterns have the advantage that all of the Fourier weight is at either DC or $k_\text{sim}$, and $I(k_\text{sim}) / I(0) = 1/3$. The mrm-SIM patterns have significant weight at intermediate frequencies (Fig.~\ref{fig:sim_fourier_analysis}b) and thus $I(k_\text{sim}) / I(0) \leq 0.1$. Therefore, for fixed intensity mrm-SIM patterns contain less signal related to the highest frequency improvements.

\begin{figure}[htb!]
	\centering
	\includegraphics[width=0.7\linewidth]{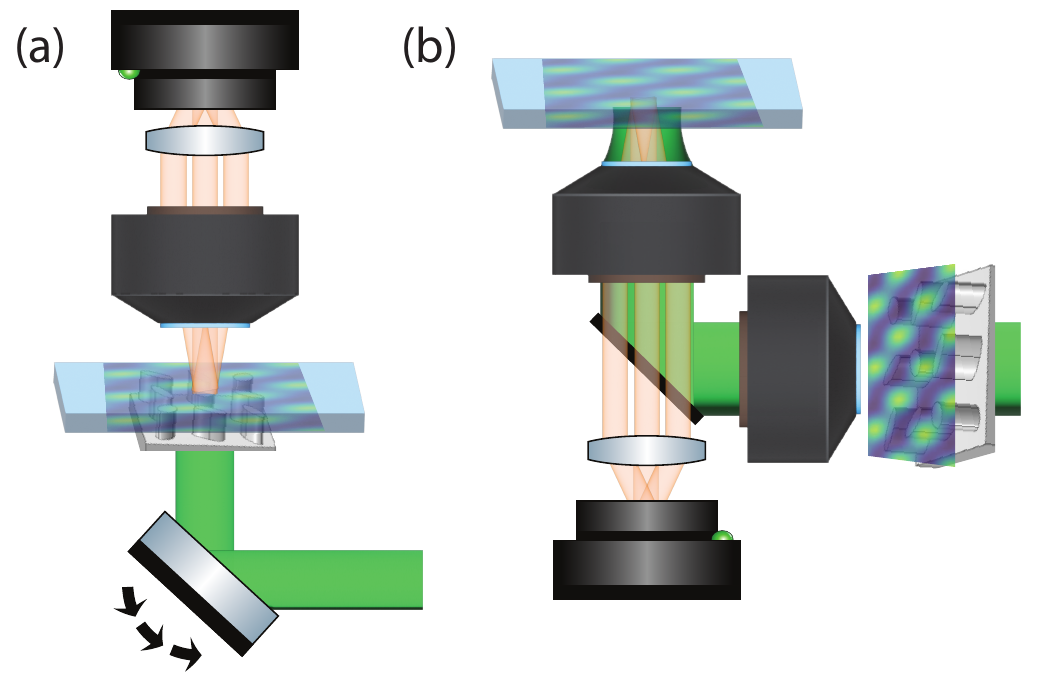}
	\caption{
    \textbf{Proposed experimental realizations}.
    (a) A proposed experimental approach to realizing mrm-SIM. The excitation light (green) impinges on the metasurface which is placed below the coverslip. Metasurface patterns excite fluorescence (orange) in a sample. The fluorescence light is collected in transmission by an imaging system and detected on a camera.
    (b) An epifluorescence microscopy approach for mrm-SIM. The metasurface is placed in a conjugate imaging plane which is optically relayed to the sample. A dichroic mirror reflects the excitation light and transmits the fluorescence.
\label{fig:sim_experiment}}
\end{figure}

Although we have so far assumed the mrm-SIM patterns are relayed to the sample, mrm-SIM is compatible with several different experimental geometries. In the simplest scheme, the metasurface could be placed immediately underneath the sample and coverslip (Fig.~\ref{fig:sim_experiment}a) and be illuminated by a collimated beam with angle control via a galvenometric mirror and polarization control via a polarization rotator or electric-optic modulator. To avoid fluorescence light interacting with the metasurface, fluorescence light should be collected in transmission and excitation light rejected with a blocking filter. Embedding the metasurface in the glass coverslip \qty{5}{\micro \meter} from the surface would result in the simplest alignment. Otherwise, the metasurface could be placed below the coverslip, in which case the evolution of the mrm-SIM patterns through \qty{\sim 100}{\micro \meter} of glass must be considered. Alternatively, if the metasurface geometry is chosen such that the excitation light is diffracted by the metasurface but, due to the Stokes shift the fluorescence light is not, such a configuration might be compatible with an epifluorescence arrangement, albeit for only a single wavelength.

mrm-SIM can be applied to an epifluorescence geometry more generally by placing the metasurface in a conjugate imaging plane and relaying the patterns to the sample, as shown in Fig.~\ref{fig:sim_experiment}b. The beam angle and polarization can be controlled similar to the approach discussed above. The epifluorescence approach makes mrm-SIM compatible with many commercial microscopes, but at the cost of introducing a more complex optical train, precise alignment, and limiting the achievable SIM pattern numerical aperture to \num{1}. 

The dependence of the mrm-SIM patterns on the lattice parameters suggests the natural way to tune their frequency is to change the lattice constant. For example, decreasing the lattice constant by $\sim 2/3$ would presumably push the SIM component at $\vec{b}_2$ to the band pass frequency. However, in our current design the metasurface already operates at nearly the maximum spatial frequency allowed for propagating the pattern in vacuum. Nevertheless, modern high-numerical aperture microscope objectives support considerably larger spatial frequencies, enabled by immersing the objective in water or oil. Immersion increases the maximum propagating spatial frequency by a factor of the refractive index of the immersion medium. Using an analogous approach, immersing the metasurface in a liquid or solid medium with $n \sim \num{1.5}$ would allow higher-frequency SIM patterns supporting larger resolution gains.

\section*{Conclusion}
In conclusion, we introduced the concept of mrm-SIM, where a wavefront-shaping, multipolar-resonant metasurface acts as an ultrathin, single surface, all-optical generator of SIM illumination patterns. The metasurface is composed of silicon nanoresonators with elliptical cross-sectional area to enable an incident angle and polarization tunable optical response. SIM-appropriate illumination patterns were achieved by leveraging geometric anisotropy of the nanoresonators, resulting in their mutually rotated assembly in a periodic supercell. An optimization algorithm was developed and employed to identify seven incident angles $\theta$ and polarization states $\varphi$ to generate SIM illumination patterns. Two types of fluorescent samples were simulated to assess the quality of mrm-SIM, and the results were directly compared to conventional hexagonal SIM. High quality image reconstructions were achieved with our mrm-SIM approach, where resolutions of \qty{240}{\nano \meter} were observed, thus reaching similar resolutions to conventional hexagonal SIM (\qty{180}{\nano \meter}). Modest resolution differences were explained by the pattern weight distribution at varying frequencies for each technique.

The proof of concept mrm-SIM demonstrated in this work operates at near-infrared (NIR) wavelengths due to the optical properties of the silicon nanoresonator material system. Such NIR illumination systems are suitable for SIM research with upconverting dyes~\cite{mettenbrinkBioimagingUpconversionNanoparticles2022,MalhotraLanthanideDopedUpconversion2023}  which, e.g., can enable new applications in deep-tissue imaging. For future work, analogous SIM-compatible metasurfaces which operate at visible frequencies can be developed with alternative material systems such as \ch{SiN_x} and \ch{TiO2} which exhibit low losses and can thus support narrow bandwidth multipolar resonances such as those presented in this work \cite{khorasaninejad2016metalenses,yang2020structural}. Furthermore, it would be advantageous to develop geometries which provide sufficient pattern diversity utilizing only polarization rotation or metasurfaces embedded in high-refractive-index media which can support SIM pattern frequencies with $\text{NA} \geq 1$. The metasurface SIM platform demonstrated here makes steps toward significant experimental simplification and democratization of SIM. The proposed all-optical tunability of the metasurface-generated far-field wavefront on a single, ultrathin metasurface paves the way toward next-generation SIM experiments with larger fields of view, decreased alignment sensitivity, miniaturized experimental footprint and minimized experimental complexity.

\section*{acknowledgement}
The authors thank Prachi Thureja and Zaid Haddadin for fruitful discussions. All authors acknowledge funding from grant number 2021- 236170 from the Chan Zuckerberg Initiative DAF, an advised fund of Silicon Valley Community Foundation. PTB and DPS acknowledge funding from Scialog, Research Corporation for Science Advancement, and Frederick Gardner Cottrell Foundation 28041. LVP acknowledges funding from the Air Force Office of Scientific Research Young Investigator Research Program Grant (FA9550-23-1-0263).

\section*{Data availability}
The electric and magnetic field data associated with figures 2--6, the synthetic image generation code, and the SIM reconstruction code are available online~\cite{associated_data}.

\section*{Supporting Information}

\renewcommand{\thesection}{S\arabic{section}}
\renewcommand{\thesubsection}{\thesection.\arabic{subsection}}
\renewcommand{\thefigure}{S\arabic{figure}} 
\renewcommand{\theequation}{S\arabic{equation}}

\setcounter{figure}{0}
\setcounter{equation}{0}

\section{Finite Element Simulations of Metasurface\label{section:FEM_simulation}}
\begin{figure}[htb!]
	\centering
	\includegraphics[width=0.45\linewidth]{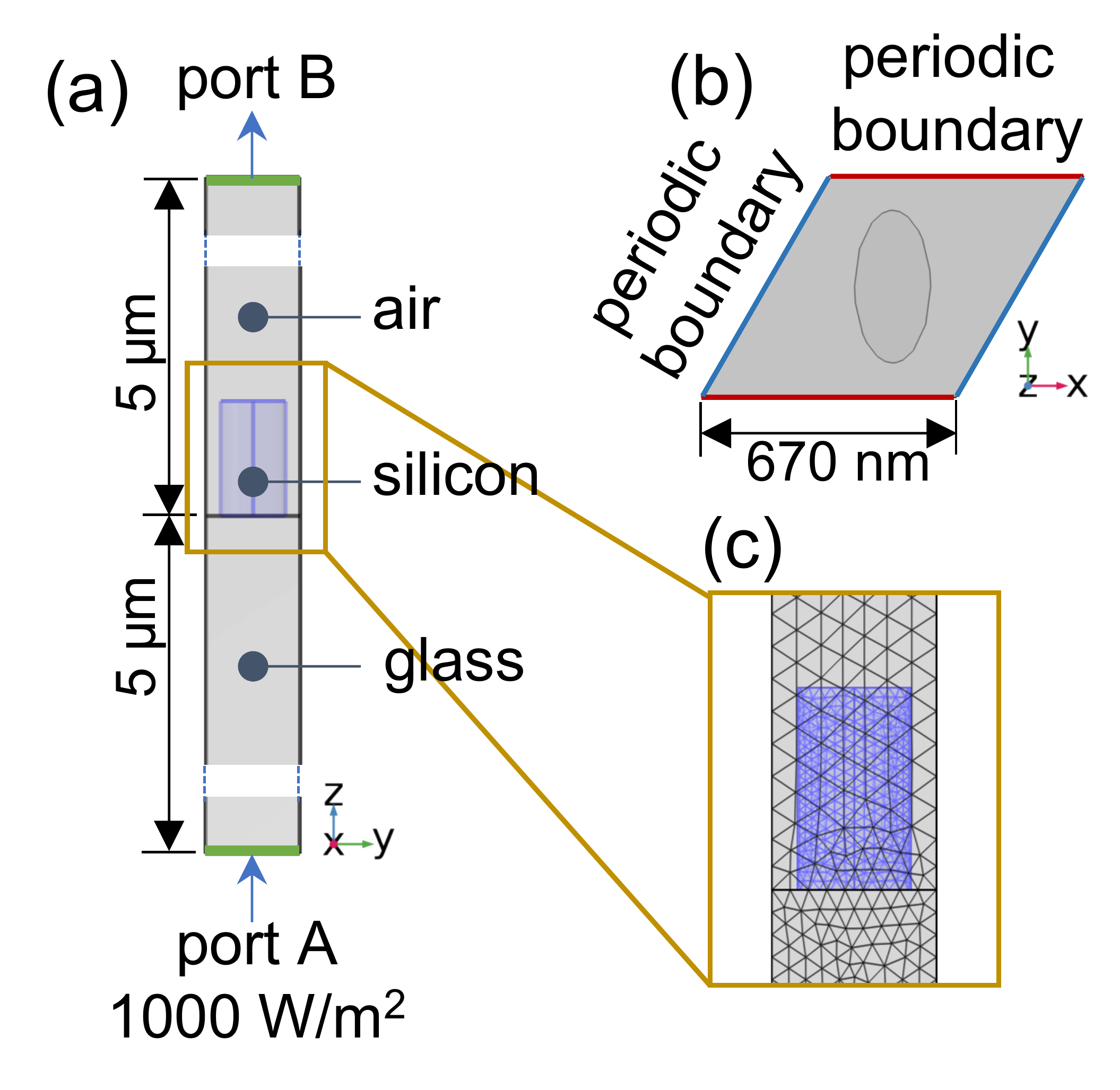}
	\caption{
    \textbf{Schematic of the simulation domain}. (a) Side view of the domain indicating the dimensions, materials, and ports. (b) Top view (z-normal) of the domain indicating the periodic boundaries. (c) Representative view of the mesh.
\label{fig:S0}}
\end{figure}

Full-field numerical computations in the frequency domain were performed with COMSOL Multiphysics 6.0, resulting in time-harmonic electromagnetic fields governed by Maxwell’s Equations. Phasor quantitites are assumed to carry time dependence $\exp (i \omega t)$, implying phasor $\exp (-i \vec{k} \cdot \vec{r})$ represents a plane wave propagating along the $\uvec{k}$ direction. The optical properties of silicon \cite{pierce1972electronic} and glass (Corning HPFS 7980 Fused Silica) were adapted from the COMSOL material library, with the refractive index of the glass and the silicon domains at the resonant wavelength of \qty{1004}{\nano \meter} as  \num{1.4508} and \num{3.6356}, respectively. The glass and the air domain spanned \qty{5}{\micro \meter} each, with silicon nanostructures on the glass-air interface 
 (Fig. \ref{fig:S0}a). 

Port boundary conditions were used on the glass and the air domains for excitation and measuring the transmitted and reflected waves. A plane wave of intensity \qty{1000}{\watt / \meter^2} was applied along the positive z-axis on the port on the glass domain. The port on the air domain was set to measure the transmitted wave. Periodic boundary conditions were applied on the side walls of the glass and the air domains (Fig. \ref{fig:S0}b). The mesh size was $\lambda/6n$ in the glass and the air domains, where $\lambda$ is the wavelength of the incident light and $n$ is the refractive index of the respective medium. The maximum element size was \qty{40}{\nano \meter} in the silicon domain (Fig. \ref{fig:S0}c). Analogous simulation parameters were employed for the mrm-SIM metasurface supercell (Fig. 2).

\section{Far Field Analysis of Illumination Patterns\label{section:far_field}}
\begin{figure}[htb!]
	\centering
	\includegraphics[width=0.7\linewidth]{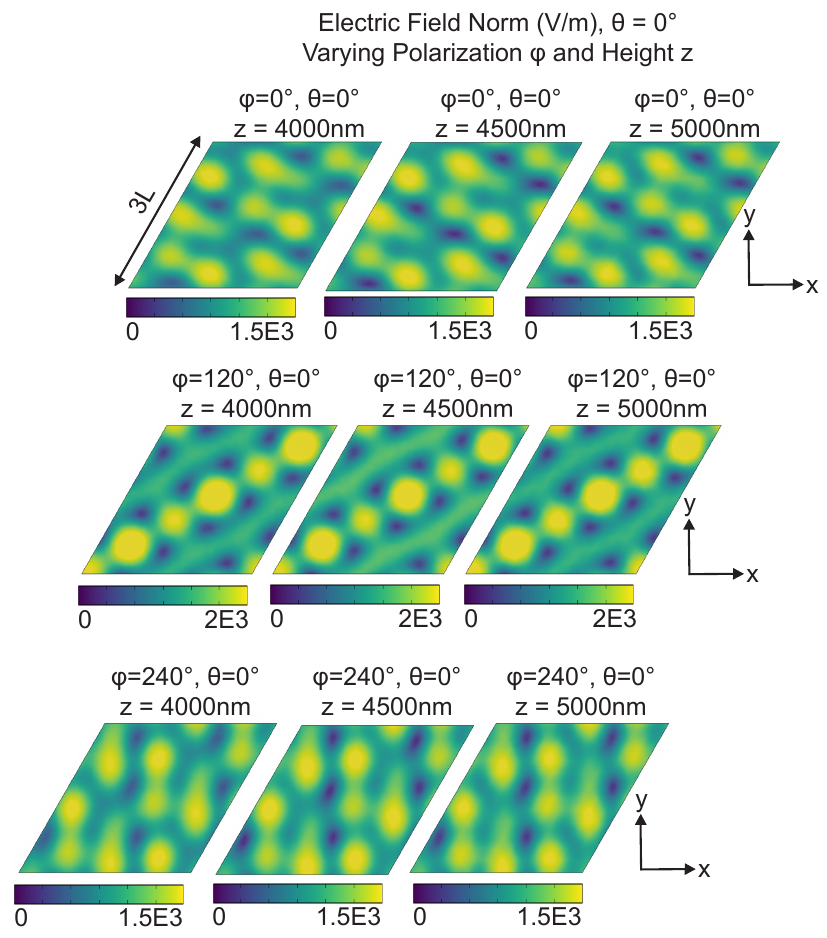}
	\caption{
    \textbf{Far-field characterization of wavefront-shaping metasurface}. Electric field norm of wavefront patterns at varying heights from $z = \{\qty{4000}{\nano \meter}, \qty{4500}{\nano \meter}, \qty{5000}{\nano \meter} \}$ (left to right) for varying incident polarizations $\varphi = \{0^\circ, 120^\circ, 240^\circ \}$ (top to bottom) and incident angle $\theta = 0^\circ$.
\label{fig:S1}}
\end{figure}

\begin{figure}[htb!]
	\centering
	\includegraphics[width=0.7\linewidth]{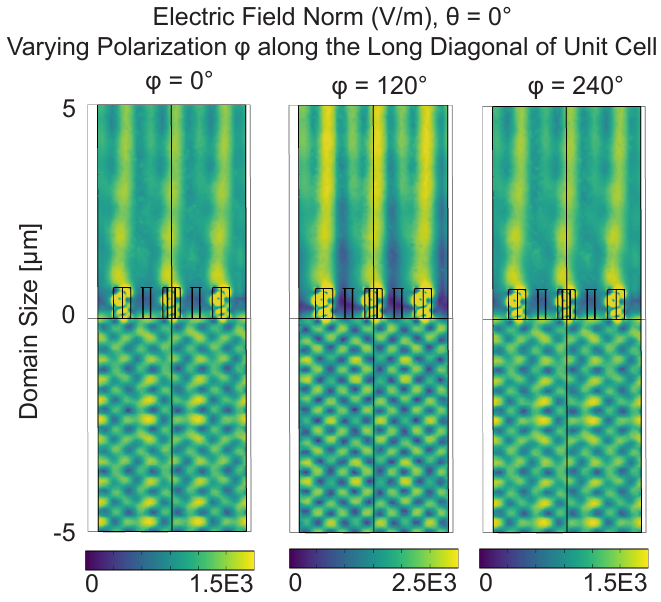}
	\caption{
    \textbf{Far-field characterization of wavefront-shaping metasurface}.
Electric field norm of wavefront patterns along the long diagonal of the metasurface supercell at varying incident polarizations $\varphi = \{0^\circ, 120^\circ, 240^\circ \}$ (left to right) for incident angle $\theta = 0^\circ$.
\label{fig:S2}}
\end{figure}

Figures \ref{fig:S1} and \ref{fig:S2} present additional information analyzing the mrm-SIM patterns in the far field. 

Figure \ref{fig:S1} shows example electric field norms of the illumination patterns generated by the multipolar resonant metasurface presented in this work for varying incident polarization states $\varphi = \{0^\circ, 120^\circ, 240^\circ\}$, on the top, middle and bottom rows, respectively. The incident angle $\theta = 0^\circ$ is studied in all cases considered. Far-field illumination patterns are shown at varying far-field heights of $z = \{\qty{4000}{\nano \meter}, \qty{4500}{\nano \meter}, \qty{5000}{\nano \meter} \}$ on the left, center and right, respectively. Fig. \ref{fig:S1} demonstrates how the illumination patterns remain consistent at these varying heights. Moreover, while the illumination patterns at $\varphi = 0^\circ$ and $\varphi = 240^\circ$ are rotations of one another, the differences in the illumination patterns for $\varphi = 120^\circ$ arise from the symmetry of the illumination source with respect to the metasurface unit cell. Because COMSOL uses a single plane wave illumination with the studied port condition, symmetry of the metasurface unit cell with respect to the incident plane wave is analogous but rotated for $\varphi = 0^\circ$ and $\varphi = 240^\circ$, while $\varphi = 120^\circ$ has a different symmetry with respect to the metasurface unit cell as the incident plane wave traverses the long axis of all three ellipses with axes along the long diagonal of the rhombohedral simulation domain. In an experiment, these differences in symmetry would average out due to a finite-sized illumination beam.

Figure \ref{fig:S2} shows the electric field norm for the studied simulation domains at $\theta = 0^\circ$ for varying incident polarizations $\varphi = \{0^\circ, 120^\circ, 240^\circ\}$ from left to right. A cross section of the simulation domain along the long diagonal of the rhombohedral unit cell shows the full z-range. This view of the electric field norm illustrates the consistence of the multipolar metasurface illumination patterns observed in the far field.

\section{Obtaining illumination patterns\label{section:interpolating_patterns}}
The electromagnetic fields obtained from the COMSOL simulations must be processed to determine the illumination intensity patterns on a square grid suitable for simulating eq.~\ref{eq:discrete_forward_model}. The COMSOL simulations are performed for a single unit cell of the metasurface, but we would like to simulate fluorescence densities covering many unit cells. To achieve this, we first duplicate the electric field patterns at coordinates shifted by $n_1 \vec{c}_1 + n_2 \vec{c}_2$ for enough $n_1, n_2$ to cover our desired field of view. Here $\vec{c}_1 = \vec{a}_1$ and $\vec{c_2} = 3 \vec{a}_2 - \vec{a}_1$ are the vectors describing the periodicity of the simulation cell. Each shifted copy of the electric field must be multiplied by a phase factor $\exp \left[-i \vec{k}_\text{in} \cdot \left( n_1 \vec{c}_1 + n_2 \vec{c}_2 \right)\right]$ to account for the Bloch (Floquet) boundary conditions used in the simulation, $\vec{k}_\text{in}$ is the wavevector of the plane wave illuminating the metasurface and $|\vec{k}_\text{in}| = 2\pi n / \lambda$ where $n$ is the refractive index of the glass. We the interpolate the electric field components onto a square grid and take the intensity pattern to be the modulus squared. This effectively assumes the fluorescence object consists of emitters with randomly oriented dipoles.

\section{Simulating realistic fluorescence images\label{section:simulating_imgs}}
We simulate realistic images using custom Python software, specifically our \texttt{localize\_psf}~\cite{localize_psf} and \texttt{mcSIM}~\cite{mcsim} packages. To simulate a realistic microscope image, we begin with a ground truth fluorophore density $S(\vec{r})$ and illumination pattern $P(\vec{r})$. Then we apply the discretized microscope forward model given in eq.~\ref{eq:discrete_forward_model}, using a vectorial model of the microscope point-spread function generated using \texttt{psfmodels}~\cite{psfmodels}. For our simulations we assumed a numerical aperture of \num{1.3}, an object grid size \numproduct{1368x1368} pixels and pixel size of \qty{29.25}{\nano \meter}, and a raw image grid pixel size of \qty{58.5}{\nano \meter}. This grid is oversampled with respect to Nyquist sampling. Next, we determine each pixel's ``noisy'' value by drawing from a Poisson distribution with mean value given by the pixel value after applying the forward model. We model the action of an sCMOS camera by assuming a gain of \qty{2}{\adu / \electron} adding an offset of \qty{100}{\adu} and Gaussian noise with standard deviation \qty{2.5}{\electron}. We do not include pixel-to-pixel variation in the camera parameters although this is expected to be present in real sCMOS images.
 
\section{Synthetic fluorescent objects\label{section:synthetic_objects}}
The filament structures are generated from a filament network file provided by Super Resolution Simulation (SuReSim)~\cite{Venkataramani2016}. The positions given in the file are shifted and scaled to match the desired field of view. The filament network file provides vertices and the filaments themselves are drawn as one-pixel wide lines. 

Synthetic line pair images are generated with spacings of \qtyrange{60}{450}{\nano \meter} in steps of \qty{30}{\nano \meter}. The line patterns are rotated by angle of \qty{15}{\degree} from the pixel grid. Each line has a flattop profile over \qty{30}{\nano \meter} which then decays exponentially to zero with decay length half a pixel. To mitigate pixelation effects, the ground truth value at each pixel is calculated by subsampling the pixel \numproduct{3x3} times.

\section{Iterative SIM reconstruction algorithm\label{section:sim_recon_supplement}}
\begin{figure}[htb!]
	\centering
	\includegraphics[width=\linewidth]{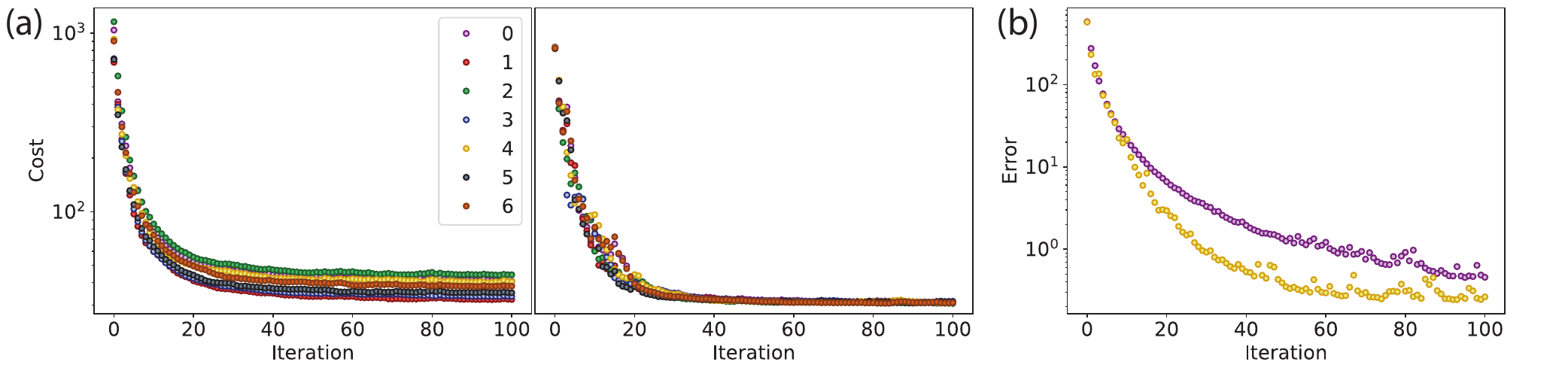}
	\caption{
    \textbf{SIM reconstruction algorithm convergence}.
	(a) The cost function evaluated for each angle. Averaging these functions produces the cost function of eq.~1 
 Results are shown for the synthetic line pairs for mrm-SIM (right panel) and hexagonal SIM (left panel) over \num{100} iterations of the algorithm.
    (b) Mean-squared error between the reconstructed fluorophore density and the ground truth values for mrm-SIM (purple) and hexagonal SIM (yellow).
\label{fig:convergence}}
\end{figure}

Suppose we have (2D) fluorophore density $S(\vec{r})$ and SIM patterns $P^{(i)}$ with ($i = 1, ... N$). The fluorescence intensity in the imaging plane is given by
\begin{align}
    I^{(i)}(\vec{r}) &=  \left[ P^{(i)}(\vec{r}) S(\vec{r}) \right] * h(\vec{r}) \label{eq:continuous_forward_model}
\end{align}
where $h$ is the point-spread function of the imaging system and $*$ is the convolution operator. 

The camera detects the integrated irradiance over each pixel, so we can rewrite this in discrete form
\begin{align}
    I^{(i)} &= B h \left[ P^{(i)} \odot S \right] \label{eq:discrete_forward_model}
\end{align}
where we regard $I$, $P^{(i)}$, and $S$ as vectors and $B$ and $h$ as linear operators. Here $\odot$ represents elementwise multiplication and $B$ is the binning operator which sums each $2 \times 2$ pixel area. Including binning allows us to define $S$ on a finer grid than the camera images, which is necessary to infer superresolution information as the camera pixel size is chosen to be Nyquist sampled with respect to the diffraction limit.

We determine the fluorophore density by solving 
\begin{align}
    S^* &= \mathop{\text{argmin}}_{S \in \mathcal{P}} \left[ \mathcal{C}(S) + \tau \text{TV}(S) \right]\label{eq:inverse_model}\\
    \mathcal{C}(S) &= \frac{1}{2N n_x n_y} \sum_i \left \Vert I^{(i)} - D^{(i)} \right \Vert^2_2
\end{align}
where $D^{(i)}$ is the measured intensity for the $i$th pattern, $\mathcal{P}$ is the set of positive valued vectors $S$, $\text{TV}$ is the total variation operator, and $\tau$ is a hyperparameter controlling the influence of the regularization term compared with the data penalty term, and $n_x n_y$ is the total number of pixels in the image $D^{(i)}$.

We solve eq.~\ref{eq:inverse_model} using the FISTA approach. Specifically, if $S_k$ is the estimated fluorophore density at the $k$th iteration, then we generate the next estimate according to
\begin{align}
    y_{k + 1} &= \text{prox} \left[ S_k - \gamma \nabla_S \mathcal{C}(S) \right] \label{eq:fista_proximal_gradient}\\
    S_{k+1} &= y_{k + 1} + \frac{q_k - 1}{q_{k + 1}} \left(y_{k+1} - y_k \right) \label{eq:fista_momentum}\\
    q_{k+1} &= \frac{1}{2} \left(1 + \sqrt{1 + 4 q_k^2} \right). \label{eq:fista_q}
\end{align}
Here we first apply gradient descent to our differentiable cost function, then incorporate the regularization terms using their proximal operators ($\text{prox}$), and finally accelerate the convergence of the scheme using a momentum type approach. We apply proximal operators to (1) project $S$ onto the space of positive vectors by setting negative values to zero and (2) enforce the TV regularization~\cite{Chambolle2004}. Here $\gamma$ is the step-size, and when $\mathcal{C}$ is convex and Lipschitz continuous with Lipschitz constant $L$ we can guarantee convergence by choosing $\gamma < 2/L$. We estimate $L$ using a power iteration scheme. We initialize the scheme by setting $S_0$ equal to the average of $D_i$ and $q_0 = 1$.

The gradient of the forward model required in eq.~\ref{eq:fista_proximal_gradient} is given by
\begin{align}
    \nabla \mathcal{C}_i &= \frac{1}{n_x n_y} M^{(i)\dag} \left(M^{(i)} - D^{(i)} \right)
\end{align}
where $M^{(i)} = B h P^{(i)}$ and $M^{(i)\dag}$ is the Hermitian adjoint of $M^{(i)}$. We find that better convergence is achieved by replacing the gradient update with a stochastic gradient update where for each iteration we randomly select a subset $\sigma_k \subseteq \{1, ..., N\}$ and replace $\nabla \mathcal{C} \to \frac{1}{|\sigma_k|} \sum_{i \in \sigma_k} \nabla \mathcal{C}_i$.

We illustrate the convergence of this algorithm for the synthetic line-pair data presented in the main text in Fig.~\ref{fig:convergence}a for \num{100} iterations. As expected for an appropriately chosen step size, the cost function is always decreasing. In Fig.~\ref{fig:convergence}b we show the mean-squared error defined by
\begin{align}
    \epsilon_k &= \frac{1}{n_x n_y} \left \Vert \left(S_k - S_\text{true}\right) * h \right \Vert^2_2.
\end{align}
We find the error decreases rapidly for both sets of SIM patterns. The error asymptotically approaches a limiting value after \num{100} iterations, and the hexagonal SIM achieves moderately smaller error than mrm-SIM due its greater high-spatial frequency content.

\section{Hexagonal SIM patterns\label{section:hex_sim}}
In the main text we compare mrm-SIM patterns with hexagonal SIM patterns. The hexagonal SIM patterns are given by
\begin{align}
    I_j(\vec{r}) &= \left \vert e^{i \vec{k}_0 \cdot \vec{r}} + e^{i \vec{k}_1 \cdot \vec{r}} e^{i \eta_j} + e^{i \vec{k}_2 \cdot \vec{r}} e^{-i \eta_j} \right \vert^2\\
    \vec{k_i} &= \frac{k_\text{sim}}{\sqrt{3}} \left( \cos \phi_i, \sin \phi_i \right)
\end{align}
with $\phi_i = \frac{2 \pi}{3} i $ for $i = 0, 1, 2$ and $\eta_j = \frac{2\pi}{7}j$ for $j = 0, ... 6$. In the main text we use $k_\text{sim} = \frac{2}{3} k_{\max} = \frac{2}{3} \frac{\text{NA}}{\lambda}$

\end{document}